\documentstyle[12pt]{article}
\newcommand{\nc}{\newcommand}  \nc{\ov}{\over} \nc{\cd}{\cdots}
\nc{\hf}{{1\ov2}} \nc{\ra}{\rightarrow} \nc{\iy}{\infty}
\nc{\bq}{\begin{equation}} \nc{\eq}{\end{equation}} \nc{\ps}{\psi}
\nc{\bZ}{{\bf Z}} \nc{\bC}{{\bf C}} \nc{\la}{\lambda} \nc{\pl}{\partial}
\nc{\inv}{^{-1}}  \nc{\al}{\alpha} \nc{\noi}{\noindent} \nc{\ph}{\varphi}
\renewcommand{\sp}{\vspace{1ex}}  \nc{\be}{\beta} \nc{\tl}{\widetilde}
\nc{\pls}{{\pl_s}} \nc{\plr}{{\pl_r}} \nc{\tr}{\mbox{tr}\ }
\nc{\m}{_{-1}} \nc{\tn}{\otimes}
\begin{document}

\begin{center}{\large\bf On the Solution of a Painlev\'e III Equation}\end{center}

\begin{center}{\bf Harold Widom\footnote{Research supported by National Science
Foundation grant DMS-9732687.}}\end{center}
\begin{center}{\it
Department of Mathematics\\ University of California, Santa Cruz, CA 95064,
USA\\ e-mail address: widom@math.ucsc.edu}\end{center}\sp

\begin{center}{\bf I.}\end{center}\sp

In a 1977 paper of McCoy, Tracy and Wu \cite{MTW} there appeared for the
first time the  solution of a Painlev\'e equation in terms of Fredholm
determinants of integral operators. Specifically, it was shown that a one-parameter family
of solutions of the equation
\bq \ps''(t)+t\inv\ps'(t)=\hf\sinh2\ps+2\al\,t\inv\sinh\ps,\label{piii}\eq
a special case of the Painlev\'e III equation, is given by
\[\ps(t)=\sum_{n=0}^{\iy}{2\ov 2n+1}\,\la^{2n+1}\int_1^{\iy}\cd\int_1^{\iy}
\prod_{j=1}^{2n+1}{e^{-tu_j}
\ov u_j+u_{j+1}}\Big[\prod_{j=1}^{2n+1}\Big({u_j-1\ov u_j+1}\Big)^{\al-\hf}\]\[+
\prod_{j=1}^{2n+1}\Big({u_j-1\ov u_j+1}\Big)^{\al+\hf}\Big]\,dy_1\cd dy_{2n+1}.\]
This is clearly expressible in terms of the Fredholm determinants of
the kernels
\[{e^{-tu}\ov u+v}\Big({u-1\ov u+1}\Big)^{\al\pm\hf}\]
acting on $L^2(1,\iy)$. The proof in \cite{MTW} is complicated, and the purpose of
this note is to give a more straightforward one.

First we give an equivalent
formulation of the solution in terms of the kernel
\[K(x,y)={e^{-t\,(x+x\inv)/2}\ov x+y}\Big|{x-1\ov x+1}\Big|^{2\al}\]
acting on $L^2(0,\iy)$. This is the representation
\[\ps=\log\,\det\,(I+{\la\ov2}\,K)-\log\,\det\,(I-{\la\ov2}\,K),\]
where $K$ is the operator with kernel $K(x,y)$. (This is very likely known but seems not to
have been writtten down in the literature before.)
To derive this second representation of $\ps$ we make the changes of variable $u_j=(x_j+x_j\inv)/2$
in the multiple integral in the first representation. Then
\[\sqrt{{u_j+1\ov u_j-1}}={x_j+1\ov x_j-1},\ \ \ du_j=\hf\,(x_j^2-1)\,{dx_j\ov x_j^2},\]
and the integral becomes
\[{1\ov 2^{2n+1}}\int_1^{\iy}\cd\int_1^{\iy}\prod_{j=1}^{2n+1}{e^{-t\,(x_j+x_j\inv)/2}
\ov (x_j+x_{j+1})\,(x_j\,x_{j+1}+1)}\Big[\prod_{j=1}^{2n+1}(x_j+1)^2\]\[+
\prod_{j=1}^{2n+1}(x_j-1)^2\Big]\Big({x_j-1\ov x_j+1}\Big)^{2\al}\Big]\,dx_1\cd dx_{2n+1}.\]
Let $f$ be an eigenfunction for $K$ with eigenvalue $\la$,
\[\int_0^{\iy}{e^{-t\,(x+x\inv)/2}\ov x+y}\Big|{x-1\ov x+1}\Big|^{2\al}\,f(y)\,dy=\la\,f(x).\]
Then the substitutions $x\ra x\inv,\ y\ra y\inv$ show that $x\inv f(x\inv)$ is also an
eigenfunction correesponding to the same eigenvalue. Hence any eigenfunction can be
written uniquely as the sum of an ``even'' eigenfunction $f_+$ satisfying
$x\inv f_+(x\inv)=f_+(x)$
and an ``odd'' eigenfunction $f_-$ satisfying $x\inv f_-(x\inv)=-f_-(x)$. The
change of variable $y\ra y\inv$ and the relations $y\inv f_{\pm}(y\inv)=\pm f_{\pm}(y)$
show that
\[\int_0^1{e^{-t\,(x+x\inv)/2}\ov x+y}\Big({1-x\ov1+x}\Big)^{2\al}\,f_{\pm}(y)\,dy=
\pm\int_1^{\iy}{e^{-t\,(x+x\inv)/2}\ov xy+1}\Big({x-1\ov x+1}\Big)^{2\al}\,f_{\pm}(y)\,dy.\]
We deduce that $f_{\pm}$ are eigenfunctions corresponding to the
eigenvalue $\la$ for the operators $K_{\pm}$ on $L^2(1,\,\iy)$ with kernels
\[K_{\pm}(x,y)=e^{-t\,(x+x\inv)/2}\,\Big[{1\ov x+y}\pm{1\ov xy+1}\Big]
\Big({x-1\ov x+1}\Big)^{2\al}
={e^{-t\,(x+x\inv)/2}\ov (x+y)\,(xy+1)}\,(x\pm1)\,(y\pm1)\Big({x-1\ov x+1}\Big)^{2\al}.\]
Hence the last displayed multiple integral equals
\[\tr K_+^{2n+1}+\tr K_-^{2n+1},\]
and it follows that
\[\psi=2\,\sum_{n=0}^{\iy}{(\hf\la)^{2n+1}\ov 2n+1}\,(\tr K_+^{2n+1}+
\tr K_-^{2n+1})=2\,\sum_{n=0}^{\iy}{(\hf\la)^{2n+1}\ov 2n+1}\,\tr K^{2n+1}\]
\[=\log\,\det\,(I+{\la\ov2}\,K)-\log\,\det\,(I-{\la\ov2}\,K),\]
as claimed.\newpage

\begin{center}{\bf II.}\end{center}\sp

Direct proofs of the fact that this function $\ps$ satisfies the
Painlev\'e equation when $\al=0$ have already been given \cite{BL,TW}. We shall make use of
some of the results of \cite{TW} here and therefore follow that paper's notation, more or
less.

First, we introduce parameters $r$ and $s$, define
\[E(x)=\sqrt{{\la\ov2}}\,e^{(rx+sx\inv)/2}\,\Big|{x-1\ov x+1}\Big|^{\al},\ \ \
K(x,y)={E(x)\,E(y)\ov x+y},\]
and let $K$ be the operator with this kernel $K(x,y)$. (In the notation of \cite{TW},
$r=t_1,\ s=t_{-1}$. The formulas
we quote from there will be in terms of our parameters $r$ and $s$.) Define
\[\ph(r,s):=\log\,\det\,(I+K)-\log\,\det\,(I-K).\]
Then $\ps(t)=\ph(-t/2,-t/2)$.  We know from \cite{TW}
that $\ph$ satisfies the sinh-Gordon equation
\bq {\pl^2\ph\ov\pl r\pl s}={1\ov2}\sinh\,2\ph.\label{sg}\eq

In order to deduce (\ref{piii}) from this we must first find a connection between the $r$
and $s$ derivatives of $\ph$. (When $\al=0$ the determinants, and so also $\ph$, depend
only on the product $rs$ and (\ref{piii}) in this case is almost immediate.) To this
end we observe that the determinants are unchanged if $K(x,y)$ is replaced
by $\tl K(x,y):=sK(sx,sy)$. This is the same as replacing $E(x)$ by
\[\tl E(x)=\sqrt{{\la\ov2}}\,e^{(rsx+x\inv)}\,\Big|{sx-1\ov sx+1}\Big|^{\al}.\]
Now
\[\pls \tl E(x)=\Big(rx+2\al {x\ov s^2x^2-1}\Big)\,\tl E(x),\]
which gives
\[\pls \tl K(x,y)=r\,\tl E(x)\,\tl E(y)+2\al {s^2xy-1\ov (s^2x^2-1)(s^2y^2-1)}
\,\tl E(x)\,\tl E(y).\]
(The denominator $x+y$ in $\tl K(x,y)$ was cancelled by its occurrence also as a factor
in both summands.) Hence
\[\pls\log\,\det\,(I+K)=\pls\log\,\det\,(I+\tl K)\]
\[=\tr (I+\tl K)\inv\Big[r\,\tl E(x)\,\tl E(y)+2\al {s^2xy-1\ov (s^2x^2-1)(s^2y^2-1)}
\,\tl E(x)\,\tl E(y)\Big].\]
(We abused notation here by writing in the bracket the kernel of the operator that is
meant.) Now we undo the variable change we made, which means we replace
$x$ by $x/s$ and $y$ by $y/s$ and divide by $s$ in the expressions for the kernels,
and we obtain
\[\pls\log\,\det\,(I+K)=\tr (I+K)\inv\Big[{r\ov s}\,E(x)\,E(y)+
{2\al\ov s} {xy-1\ov (x^2-1)(y^2-1)}\,E(x)\,E(y)\Big].\]
If we had differentiated with respect to $r$ without making a preliminary variable change
we would have obtained
\[\plr\log\,\det\,(I+K)=\tr (I+K)\inv\,E(x)\,E(y).\]
Hence we have shown that
\[s\,\pls\log\,\det\,(I+K)-r\,\plr\log\,\det\,(I+K)=2\al\,\tr (I+K)\inv
\Big[{xy-1\ov (x^2-1)(y^2-1)}\,E(x)\,E(y)\Big].\]
Replacing $K$ by $-K$ and subtracting gives the relation (we use subscript notation
for derivatives)
\bq r\,\ph_r-s\,\ph_s=4\al\,\tr (I-K^2)\inv \Big[{xy-1\ov (x^2-1)(y^2-1)}\,E(x)\,E(y)\Big].
\label{ders}\eq
This is the desired connection between the $r$ and $s$ derivatives of $\ph$.\sp

\begin{center}{\bf III.}\end{center}\sp

We want to show that $\psi(t)=\ph(-t/2,-t/2)$ satisfies (\ref{piii}), but because of
those awkward factors $-1/2$ we prefer to derive the equivalent equation
\bq {d^2\ov dt^2}\ph(t,t)+t\inv{d\ov dt}\ph(t,t)=2\,\sinh 2\ph(t,t)-4\al\,t\inv
\sinh \ph(t,t).
\label{piii2}\eq
We are going to use
\bq {d^2\ov dt^2}\ph(t,t)=2\ph_{rs}(t,t)+\ph_{rr}(t,t)+\ph_{ss}(t,t),\ \ \
{d\ov dt}\ph(t,t)=\ph_r(t,t)+\ph_s(t,t).\label{phders}\eq
Now we know that $\ph(r,s)$ satisfies the sinh-Gordon equation (\ref{sg}) so let us
see what identity we have to derive. Set
\[T=\tr (I-K^2)\inv\Big[ {xy-1\ov (x^2-1)(y^2-1)}\,E(x)\,E(y)\Big].\]
Differentiating (\ref{ders}) with respect to $r$ and $s$ gives
\[r\,\ph_{rr}+\ph_r-s\,\ph_{rs}=4\al\,T_r,\ \ \ r\,\ph_{rs}-s\,\ph_{ss}-\ph_s=
4\al\,T_s.\]
Therefore
\[-(r+s)\,\ph_{rs}+r\,\ph_{rr}+s\,\ph_{ss}+\ph_r+\ph_s=4\al\,(T_r-T_s),\]
\[(r+s)\,\ph_{rs}+r\,\ph_{rr}+s\,\ph_{ss}+\ph_r+\ph_s=2\,(r+s)\,\ph_{rs}+
4\al\,(T_r-T_s).\]
Setting $r=s=t$ and using (\ref{phders}) we get
\[t{d^2\ov dt^2}\ph(t,t)+{d\ov dt}\ph(t,t)=4t\,\ph_{rs}(t,t)+4\al\,(T_r-T_s)(t,t).\]
Hence, by  (\ref{sg}),
\[{d^2\ov dt^2}\ph(t,t)+t^{-1}{d\ov dt}\ph(t,t)=2\,\sinh2\,\ph(t,t)+
4\al\,t^{-1}(T_r-T_s)(t,t).\]
It follows that (\ref{piii2}) is equivalent to
\[(T_r-T_s)(t,t)=-\sinh\ph(t,t).\]

\begin{center}{\bf IV.}\end{center}\sp

The functions
\[E_i(x)=x^i\,E(x),\ \ F_i(x)={E_i(x)\ov x^2-1},\ \ Q_i=(I-K^2)\inv E_i,\ \
P_i=(I-K^2)\inv KE_i\]
will arise in the computations leading to this identity. We have
\[T_r-T_s=\tr (I-K^2)\inv\Big[{xy-1\ov (x^2-1)(y^2-1)}\,(\plr-\pls)\,
E(x)\,E(y)\Big]\]
\bq+\tr (\plr-\pls)(I-K^2)\inv\,\Big[{xy-1\ov (x^2-1)(y^2-1)}\,E(x)\,E(y)\Big].
\label{TrTs}\eq
Now
\[(\plr-\pls)\,E(x)\,E(y)=\hf(x-x\inv+y-y\inv)\,E(x)\,E(y),\]
which gives
\[{xy-1\ov (x^2-1)(y^2-1)}(\plr-\pls)\,E(x)\,E(y)=
\hf\Big({y-x\inv\ov y^2-1}+{x-y\inv\ov x^2-1}\Big)\,E(x)\,E(y).\]
Hence the first summand in (\ref{TrTs}) equals
\bq (Q_0,\,F_1)-(Q\m,\,F_0).\label{s1}\eq
Next, using the notation $a\tn b$ for the operator with kernel $a(x)\,b(y)$,
we have (\cite{TW}, p. 4)
\[(\plr-\pls)(I-K^2)\inv=\hf(P_0\tn Q_0+Q_0\tn P_0-P\m\tn Q\m+Q\m\tn P\m),\]
and it follows that the second summand in (\ref{TrTs}) equals
\bq (Q_0,\,F_1)\,(P_0,\,F_1)-(Q_0,\,F_0)\,(P_0,\,F_0)-(Q\m,\,F_1)\,(P\m,\,F_1)+
(Q\m,\,F_0)\,(P\m,\,F_0).\label{s2}\eq

We introduce notations for the various inner products:
\[U_{i,j}=(Q_i,\,F_j),\ \ \ V_{i,j}=(P_i,\,F_j).\]
These are analogous to the inner products
\[u_{i,j}=(Q_i,\,E_j),\ \ \ v_{i,j}=(P_i,\,E_j)\]
which play a crucial role in \cite{TW} and will here, also.
We have shown that
\bq T_r-T_s=U_{0,1}-U_{-1,0}+U_{0,1}\,U_{0,1}-U_{0,0}\,V_{0,0}-U_{-1,1}\,V_{-1,1}+
U_{-1,0}\,V_{-1,0}.\label{Trs}\eq

There are relations among the various quantities appearing here. If we set
\[u_i=u_{0,i},\ \ \ v_i=v_{0,i},\ \ \ U_i=U_{0,i},\ \ \ V_i=V_{0,i}\]
then we have the recursion formulas
\[x\,Q_i(x)-Q_{i+1}(x)=v_i\,Q_0(x)-u_i\,P_0(x),\]
\[x\,P_i(x)+P_{i+1}(x)=u_i\,Q_0(x)-v_i\,P_0(x).\]
(The first is formula (9) of \cite{TW}, the second is obtained similarly.)
Taking inner products with $E_j$ gives the formulas
\bq u_{i,j+1}-u_{i+1,j}=v_i\,u_j-u_i\,v_{j},\ \ \
v_{i,j+1}+v_{i+1,j}=u_i\,u_j-v_i\,v_{j}\label{uv}\eq
of \cite{TW} and taking inner products with $F_j$ gives the analogous formulas
\bq U_{i,j+1}-U_{i+1,j}=v_i\,U_j-u_i\,V_j,\ \ \
V_{i,j+1}+V_{i+1,j}=u_i\,U_j-v_i\,V_j.\label{UV}\eq
Observe the special case $i=j=-1$ of the second part of (\ref{uv}):
\bq u\m^2+1=(1+v\m)^2.\label{uvm}\eq
(The $u_{i,j}$ are symmetric in  $i$ and $j$.) In fact (\cite{TW}, p. 8),
\[u\m=\sinh\ph,\ \ \ 1+v\m=\cosh\ph.\]

We see from the above formulas that all the $U_{i,j}$ and $V_{i,j}$ may be expressed
in terms of the $U_i$ and $V_i$ (with coefficients involving the $u_i$ and $v_i$).
But notice that $F_{i+2}-F_i=E_i$
(here we use the form of $F_i$ for the first time). This gives $U_{i+2}-U_i=u_i,\
V_{i+2}-V_i=v_i$ and using this also it is clear that everything
can be expressed in terms of the four unknown quantities
$U_0,\; V_0,\; U_1$ and $V_1$ (and the $u_i$ and $v_i$). Using (\ref{uvm}) also
we compute that (\ref{Trs}) equals
\[ -v\m\,U_1+u\m\,V_1+u\m\]
\bq +U_1\,V_1-U_0\,V_0-((1+v\m)\,U_0-u\m\,V_0)\,(u\m\,U_0-(1+v\m)\,V_0)\label{Trs1}\eq
\[ +((1+v\m)\,U_1-u\m\,V_1-u\m)\,(u\m\,U_1-(1+v\m)\,V_1-v\m).\]

 Now we are going to use, as we did before, the fact that conjugation by the unitary
 operator $f(x)\ra x\inv f(x\inv)$ has the effect on $K$ of interchanging $r$ and $s$.
 Thus $K$ is invariant under this conjugation when $r=s$. Since $E_i$ is sent to
 $E_{-i-1}$ and $F_i$ to $-F_{-i+1}$ we find that when $r=s$
 \[U_0=-(Q\m,\,F\m)=-(1+v\m)\,U_0+u\m\,V_0,\]
 \[U_1=-(Q\m,\,F_0)=-(1+v\m)\,U\m+u\m\,V\m=-(1+v\m)\,(U_1-u\m)+u\m\,(V_1-v\m).\]
 From these we deduce that
 \bq V_0={u\m\ov v\m}\,U_0,\ \ \ V_1={u\m\ov v\m}\,U_1-1.\label{V}\eq
 Using these we find that when $r=s$ (\ref{Trs1}) simplifies to
 \[2{u\m\ov v\m}(U_0^2-U_1^2).\]
 Since $u\m=\sinh\ph$ we have reduced the problem to showing that
 \bq U_1(t,t)^2-U_0(t,t)^2=\hf\,v\m(t,t).\label{UV1}\eq

\begin{center}{\bf V.}\end{center}\sp

Let us compute $d/dt$ of both sides of the desired identity.
Of course, $d/dt\,U_0(t,t)=(\plr+\pls)\,U_0(t,t)$, etc.,  so we begin by writing down these
derivatives. We have
\[2\,(\plr+\pls)\,E_i=E_{i+1}+E_{i-1},\ \ \ 2\,(\plr+\pls)\,F_i=F_{i+1}+F_{i-1},\]
\[2\,(\plr+\pls)\,(I-K^2)\inv=\hf(P_0\tn Q_0+Q_0\tn P_0+P\m\tn Q\m+Q\m\tn P\m).\]
(For the last, see \cite{TW}, p. 4.) Using these we compute
\[2\,(\plr+\pls)\,U_i=2\,U_{i+1}+2\,u_0\,V_i+(1+u\m^2+(1+v\m)^2)\,U_{i-1}-
2\,u\m\,(1+v\m)^2\,V_{i-1}.\]
Taking $i=0$ and 1 and using (\ref{V}) and (\ref{uvm}) we find that
when $r=s$
\[2\,(\plr+\pls)\,U_0=2\,u_0\,{u\m\ov v\m}\,U_0-2\,v\m\,U_1,\]
\[2\,(\plr+\pls)\,U_1=2\,u_0\,{u\m\ov v\m}\,U_1-2\,v\m\,U_0.\]
Hence
\bq (\plr+\pls)\,(U_1^2-U_0^2)=2\,u_0\,{u\m\ov v\m}\,(U_1^2-U_0^2).\label{Uder}\eq

Now we compute in a similar way (cf., \cite{TW}, p. 5)
\[2\,(\plr+\pls)\,v_i=u\m\,u_0+v\m\,v_0+v_0+v_{-1,1}+u_{-1,-1}\,u\m+v_{-1,-1}\,v\m
+v_{-2}+v_{-1,-1}.\]
Again applying the operator $f(x)\ra x\inv f(x\inv)$ we find that when $r=s$ we have
\[u_{-1,-1}=u_0,\ \ \ v_{-1,-1}=v_0,\ \ \ v_{-2}=v_{-1,1},\]
so the above is
\[2\,(u\m\,u_0+v\m\,v_0+v_0+v_{-1,1}).\]
Applying the second part of (\ref{uv}) with $i=-1,\;j=0$ gives
$v_0+v_{-1,1}=u\m\,u_0-v\m\,v_0$, and so we have shown that when $r=s$
\[(\plr+\pls)\,v\m=2\,u\m\,u_0=2\,u_0\,{u\m\ov v\m}\,v\m.\]
This relation and (\ref{Uder}) show that $U_1(t,t)^2-U_0(t,t)^2$ and $v\m(t,t)$ are equal up
to a constant factor, and to deduce (\ref{UV1}) it remains only to compute this factor.
We do this by determining the asymptotics of both quantities as $t\ra-\iy$. For convenience
we evaluate everything at $r=s=-t$ and let $t\ra+\iy$.

We have
\[v\m=\Big((I-K^2)\inv K\,E,\,E\m\Big).\]
If we were to replace $(I-K^2)\inv$ by $I$ we would be left with
\[(KE,\,E\m)=\int_0^{\iy}\int_0^{\iy}{E(x)^2\,E(y)^2\ov x+y}y\inv\,dy\,dx
\sim \hf\Big(\int_0^{\iy}E(x)^2\,dx\Big)^2\]
since the main contributions to the integrals come from neighborhoods of $x=y=1$.
It is an easy exercise to show that
\bq\int_0^{\iy}e^{-t\,(x+x\inv)}\Big|{x-1\ov x+1}\Big|^{2\al}\,dx
\sim \Gamma(\al+\hf)\,2^{-2\al}\,t^{-\al-\hf}\,e^{-2t},\label{EE}\eq
and so
\[(KE,\,E\m)\sim {\la^2\ov 4}\,\Gamma(\al+\hf)^2\,2^{-4\al-1}\,t^{-2\al-1}\,e^{-4t}.\]
The error caused by our replacement of $(I-K^2)\inv$ by $I$ is of smaller order
of magnitute. This follows from the fact that the square of the $L^2$ norm of $E$
is $O(t^{-\al-\hf}\,e^{-2t})$, as shown above, and hence so is the operator norm of $K$.
Thus the error, which equals $((I-K^2)\inv K^3\,E,\,E\m)$, is $O(t^{-4\al-2}\,e^{-8t})$.
Therefore we have shown
\[v\m\sim {\la^2\ov 4}\,\Gamma(\al+\hf)^2\,2^{-4\al-1}\,t^{-2\al-1}\,e^{-4t}.\]

Next,
\[U_1-U_0=\Big((I-K^2)\inv E,\,(x+1)\inv E\Big),\ \ \ U_1+U_0=
\Big((I-K^2)\inv E,\,(x-1)\inv E\Big)\]
and $U_1^2-U_0^2$ is the product of these. As before, replacing $(I-K^2)\inv$ by $I$
in each factor will not affect the first-order asymptotics of the product.
After this replacement the first inner product becomes $\la/2$ times the integral in
(\ref{EE}) but with an extra factor $x+1$ in the denominator. Thus
\[U_1-U_0\sim {\la\ov2}\,\Gamma(\al+\hf)\,2^{-2\al-1}\,t^{-\al-\hf}\,e^{-2t}.\]
After the replacement the second inner product becomes $\la/2$ times
\[\int_0^{\iy}e^{-t\,(x+x\inv)}\Big|{x-1\ov x+1}\Big|^{2\al}\,{dx\ov x-1}.\]
This is a little trickier since when we make the variable change $x=1+y$ to compute
the asympotics we must use the second-order approximations $x+x\inv\ra 2+y^2-y^3$
and $(x+1)^{-2\al}\ra 2^{-2\al}(1-\al y)$. But it is still straightforward and
we find that
\[U_1+U_0\sim {\la\ov2}\,\Gamma(\al+\hf)\,2^{-2\al-1}\,t^{-\al-\hf}\,e^{-2t}.\]
Thus
\[U_1^2-U_0^2\sim {\la^2\ov4}\,\Gamma(\al+\hf)^2\,2^{-4\al-2}\,t^{-2\al-1}\,e^{-4t}
\sim \hf\,v\m.\]

We knew that $(U_1^2-U_0^2)/v\m$ is a constant and now we see that the constant equals
$1/2$. This establishes (\ref{UV1}) and concludes the proof.

\end{document}